\documentclass[twocolumn,aps,prd,10pt,superscriptaddress,longbibliography]{revtex4-1}

\usepackage{placeins}

\usepackage{amssymb,amsmath}
\usepackage{bm}
\usepackage{dcolumn}% Align table columns on decimal point
\usepackage{glossaries}
\usepackage{graphicx}% Include figure files
\usepackage[version=4]{mhchem}
\usepackage[usenames,dvipsnames,svgnames]{xcolor}
\usepackage{siunitx}
\usepackage{hyperref}
\usepackage{makecell}
\hypersetup{
pdfnewwindow=true,      % links in new window
colorlinks=true,        % false: boxed links; true: colored links
linkcolor=Blue,         % color of internal links
citecolor=Blue,         % color of links to bibliography
filecolor=Blue,         % color of file links
urlcolor=Blue           % color of external links
}

\newacronym{mof}{MOF}{metal-organic framework}
\newacronym{nep}{NEP}{neuroevolution potential}
\newacronym{sm}{SM}{Supplemental Material}
\newacronym{md}{MD}{molecular dynamics}
\newacronym{pimd}{PIMD}{path-integral molecular dynamics}
\newacronym{mlp}{MLP}{machine-learned potential}
\newacronym{nte}{NTE}{negative thermal expansion}
\newacronym{pte}{PTE}{positive thermal expansion}
\newacronym{spc}{SPC}{soft porous crystal}
\newacronym{tec}{TEC}{thermal expansion coefficient}
\newacronym{nqe}{NQE}{nuclear quantum effect}
\newacronym{vdos}{VDOS}{vibrational density of states}

\DeclareSIUnit\angstrom{\text{Å}}
\DeclareSIUnit{\atom}{atom}
\DeclareSIUnit{\step}{step}
\DeclareSIUnit{\atomstepsecond}{\atom\step\per\second}

\begin{document}
\title{Side-Chain Tuning of Thermal-Expansion Crossover in Metal-Organic Frameworks}

\author{Wei Qiu}
% \email{weiqiu@tauex.tau.ac.il}
\affiliation{Department of Physical Chemistry, School of Chemistry, The Raymond and Beverly Sackler Faculty of Exact Sciences and
The Sackler Center for Computational Molecular and Materials Science, Tel Aviv University, Tel Aviv 6997801, Israel}

\author{Penghua Ying}
\email{penghua@xjtu.edu.cn}
\affiliation{Laboratory for multiscale mechanics and medical science, SV LAB, School of Aerospace, Xi’an Jiaotong University, Xi’an, Shaanxi, 710049, China}

\date{\today}

\begin{abstract}

Achieving continuous control over macroscopic thermal expansion remains a fundamental challenge in solid-state physics. Using classical and path-integral molecular dynamics alongside lattice dynamics at near-\emph{ab initio} accuracy, we report an entropy-driven thermal-expansion crossover from positive (PTE) to negative thermal expansion (NTE) in alkoxy-functionalized MOF-5, an archetypal metal-organic framework (MOF). We demonstrate that this non-linear response is continuously tunable via the alkoxy side-chain length, quantified by the number of carbon atoms $n$ grafted onto the archetypal cubic MOF-5 framework: systems with short chains ($n \le 2$) exhibit monotonic NTE, whereas longer chains ($n \ge 3$) trigger a pronounced PTE-to-NTE crossover. At low temperatures, thermal activation of longer side chains opens additional conformational states and generates steric pressure inside the pore, driving positive expansion through a gain in side-chain conformational entropy. Conversely, at elevated temperatures, the side chains enhance transverse linker fluctuations and strengthen the string-tension mechanism associated with low-frequency framework modes, causing structural contraction favored by framework vibrational entropy. Finally, by varying the concentration of side-chain-functionalized linkers, the thermal expansion coefficient can be continuously regulated to realize negative, near-zero, and positive thermal expansion within selected temperature windows. These results establish side-chain engineering as a practical route for programming macroscopic thermodynamic responses in MOFs.

\end{abstract}
\maketitle
\raggedbottom

Thermal expansion is a fundamental thermodynamic property that reflects the underlying lattice dynamics and anharmonicity of solids~\cite{gruneisen1912theorie}. While most materials expand upon heating, anomalous \gls{nte} provides a route to zero-expansion composites and a platform for exploring unusual phonon physics~\cite{mary1996negative, evans1996negative, greve2010pronounced, lightfoot1998widespread, shaikh2025negative, tian2024machine, li2022chemical,wu2008negative,liu2026isotropic,chen2026unified,yang2026structural}. Among various \gls{nte}-exhibiting materials, \glspl{mof} stand out because of their exceptional structural designability and chemical diversity~\cite{yaghi2003reticular}. The inherent flexibility of these \glspl{mof} allows for a rich variety of low-frequency vibrational modes, including transverse ``guitar-string'' vibrations of organic linkers~\cite{zhou2008origin}, which can shorten the projected linker length and drive macroscopic contraction.

Precise control of the \gls{tec} in \glspl{mof} and related framework materials is crucial for mitigating thermal stress and maintaining dimensional stability in precision instruments, optical systems, microelectronic packaging, and composite structures, where thermal-expansion mismatch can lead to deformation, interfacial failure, or performance drift~\cite{takenaka2012negative,li2022chemical,mary1996negative,sleight1998isotropic,van1999origin}. However, achieving such control remains a formidable challenge. Current design strategies often focus on mixed-ligand systems~\cite{sethi2023tuning,baxter2019tuning}, topological design~\cite{burtch2019negative}, guest adsorption~\cite{balestra2016controlling}, or additional linkages that stiffen the parent framework~\cite{schneider2019tuning}. These approaches have successfully tuned the magnitude of thermal expansion, but they are usually based on static structural or compositional modifications. As a result, they rarely produce a temperature-driven sign reversal within a single empty framework. Recent studies of superionic solids further demonstrate that mobile internal degrees of freedom can strongly reshape the thermodynamic response of crystalline hosts~\cite{nagaya2026experimental}. This raises a broader question: can internal molecular degrees of freedom be used as chemical handles to program not only the magnitude, but also the sign, of thermal expansion in MOFs?

To address this question, we use alkoxy-functionalized MOF-5 as an archetypal platform in which internal molecular degrees of freedom can be systematically varied by changing the side-chain length. \Gls{md} simulations have emerged as an indispensable tool for decoding the microscopic origins of complex thermomechanical responses, enabling the assessment and prediction of \gls{nte} in complex \gls{mof} crystals~\cite{evans2019assessing}. Accurately modeling these flexible frameworks requires a rigorous treatment of both strong anharmonicity and \glspl{nqe}, necessitating advanced simulation approaches driven by potential models at near-\emph{ab initio} accuracy~\cite{feynman1965quantum, parrinello1984study, markland2018nuclear, ying2025highly,ceriotti2010efficient,peterson2010local}. In this Letter, we combine classical \gls{md}, \gls{pimd}, and lattice-dynamical analyses using a dedicated \gls{mlp}~\cite{ying2026structurally, fan2021neuroevolution} to investigate MOF-5 architectures decorated with alkoxy side chains of varying lengths~\cite{pallach2021frustrated}. We show that increasing the alkoxy side-chain length tunes MOF-5 from monotonic \gls{nte} to a pronounced positive-to-negative thermal-expansion crossover. Rather than a simple steric perturbation, the side chains introduce competing entropic contributions: side-chain conformational entropy favors low-temperature expansion, whereas low-frequency framework vibrational entropy favors high-temperature contraction. By revealing how side-chain dynamics enhance transverse linker fluctuations and the associated string-tension mechanism, we show that microscopic side-chain engineering can program the macroscopic \gls{tec}, including negative, near-zero, and positive values within selected temperature windows.

\begin{figure}[h!]
\centering
\includegraphics[width=1.0\columnwidth]{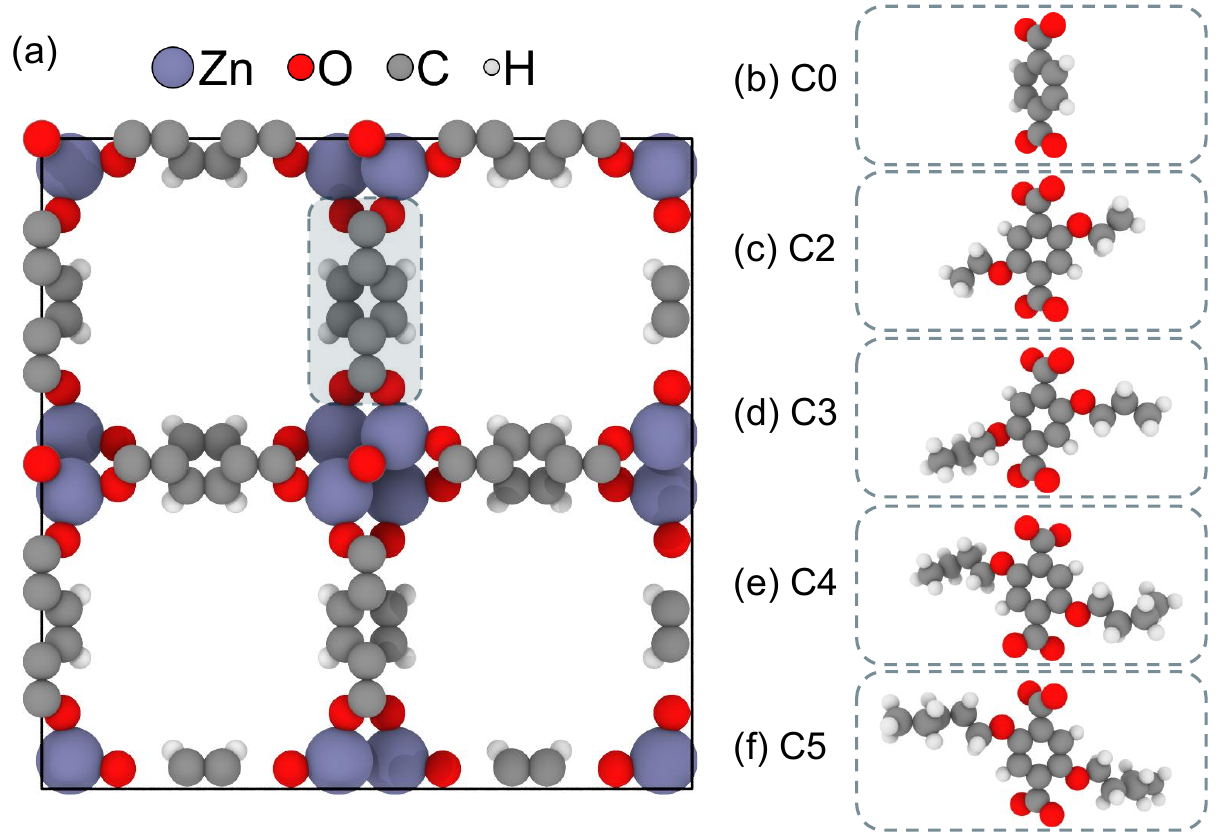}
\caption{Structural models of pristine and alkoxy-functionalized \gls{mof}-5. (a) Pristine MOF-5 framework, with the shaded region highlighting the organic linker backbone. (b)--(f) Local linker structures of C0, C2, C3, C4, and C5, where C$n$ denotes an alkoxy side chain containing $n$ carbon atoms. Color scheme: Zn, purple; O, red; C, gray; H, white. Structures were visualized using \textsc{ovito} package~\cite{stukowski2010visualization}.}
\label{fig:structure}
\end{figure}

All \gls{md} simulations were performed using the \textsc{gpumd} package~\cite{xu2025gpumd}, with a machine-learned \gls{nep}+D3 model~\cite{fan2021neuroevolution,ying2024combining} developed in Ref.~\cite{ying2026structurally}. In this model, the \gls{nep} describes short-range interatomic interactions, whereas the D3 term~\cite{grimme2010jcp} accounts for long-range dispersion interactions~\cite{ying2024combining} (see \gls{sm} Sec.~1 and Sec.~2 for details). The \gls{nep}+D3 framework has been validated for the thermal-transport properties of pristine MOF-5~\cite{li1999design} and experimentally realized alkoxy-functionalized derivatives~\cite{pallach2021frustrated}, demonstrating its ability to capture the strong anharmonicity of these flexible frameworks~\cite{ying2026structurally}. We systematically investigated a series of MOF-5-based frameworks, denoted as C$n$, where $n$ is the number of carbon atoms in the alkoxy side chain. The series includes pristine MOF-5 (C0) and four functionalized frameworks with linear alkoxy side chains grafted onto the primary organic linkers: ethoxy (C2), propoxy (C3), butoxy (C4), and pentoxy (C5), with atomic structures shown in \autoref{fig:structure}. Because finite-size effects can influence the equilibrium volume and the apparent thermal-expansion behavior~\cite{ying2025highly}, we performed size-convergence tests using \(1\times1\times1\) to \(5\times5\times5\) supercells constructed from the conventional cubic unit cell (\autoref{fig:structure}(a)). The resulting volume-temperature curves become well converged for the \(5\times5\times5\) supercells (\gls{sm} Figure S2); therefore, unless otherwise stated, all thermal expansion simulations reported below were performed using \(5\times5\times5\) supercells.

\begin{figure}[htb]
\begin{center}
\includegraphics[width=\columnwidth]{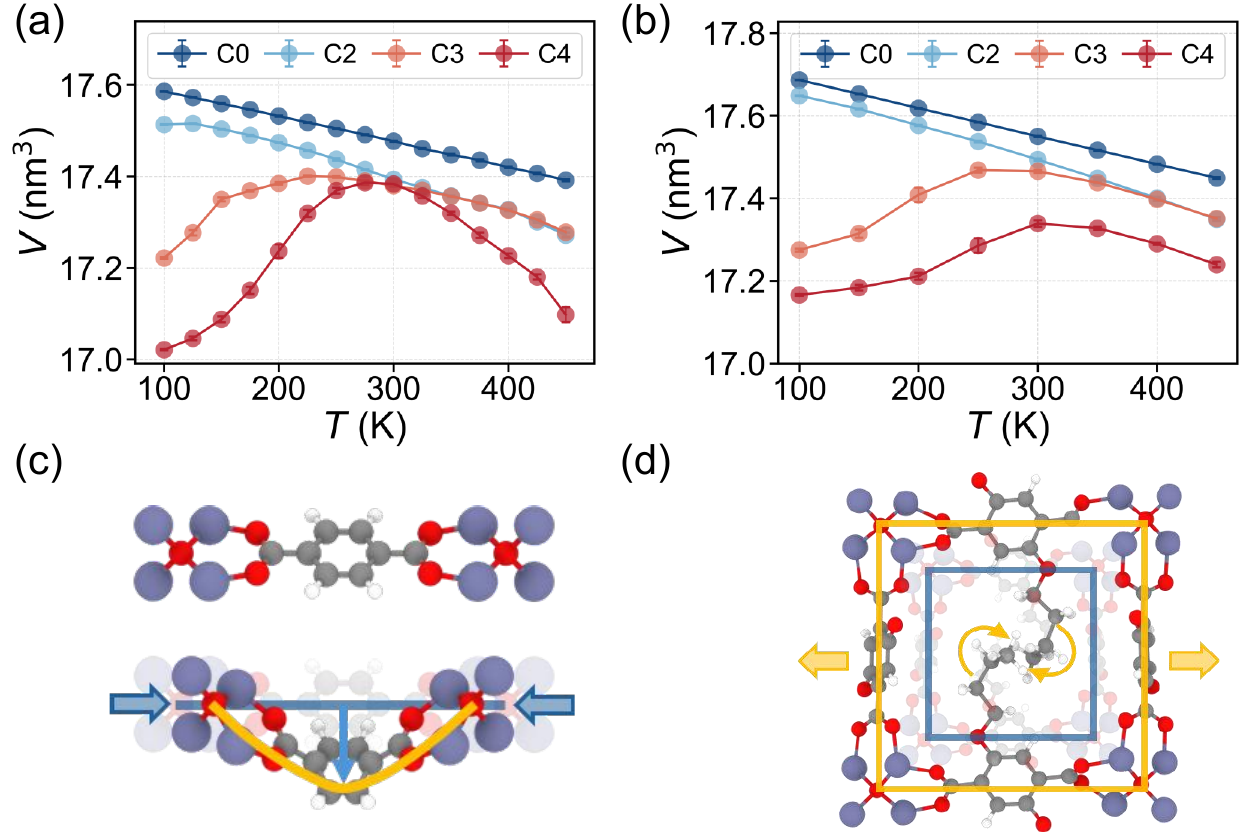}
\caption{Side-chain tuning of thermal-expansion crossover in alkoxy-functionalized \gls{mof}-5. (a,b) Temperature-dependent equilibrium volumes per unit cell for C0--C4 from (a) classical \gls{md} and (b) \gls{pimd} simulations using \(5\times5\times5\) supercells. (c) String-tension mechanism in pristine \gls{mof}-5: the blue line denotes the low-temperature linker backbone and the orange curve illustrates thermally activated transverse swinging, which shortens the projected linker length and drives \gls{nte}. (d) Side-chain opening in functionalized \gls{mof}-5: the blue square denotes the low-temperature pore outline, the orange square denotes the expanded pore after side-chain opening, and the orange arrows indicate side-chain rotation and the resulting framework expansion that promotes low-temperature \gls{pte}.}
\label{fig:crossover}
\end{center}
\end{figure}

With this C$n$ series, we examine how the equilibrium volume evolves with temperature as the side-chain length increases. Classical \gls{md} was used to determine the thermal response, while \gls{pimd} simulations were performed to explicitly include \glspl{nqe} and assess their influence on the thermal-expansion behavior. In \gls{pimd}, each nucleus is mapped onto a ring polymer composed of multiple beads, allowing \glspl{nqe} to be included in finite-temperature sampling using the stochastic \gls{pimd} thermostat~\cite{ceriotti2010efficient} implemented in \textsc{gpumd}~\cite{ying2025highly}. Simulation details for classical \gls{md} and \gls{pimd} are provided in \gls{sm} Secs. 1 and 2, respectively, and the bead-number convergence of the calculated volume-temperature behavior is documented in \gls{sm} Figure S1. This comparison reveals the central thermal expansion phenomenology of the C$n$ series, as summarized in \autoref{fig:crossover}.

As shown by the classical \gls{md} simulations in~\autoref{fig:crossover}(a), pristine MOF-5 (C0) exhibits monotonic volume contraction with increasing temperature, consistent with its well-known \gls{nte} behavior~\cite{zhou2008origin}. C2 behaves similarly, indicating that a short ethoxy side chain does not overcome the intrinsic contraction tendency of the framework. By contrast, C3 and C4 display a pronounced non-monotonic volume evolution: the volume first increases upon heating, reaches a maximum at intermediate temperature (\SI{225}{\kelvin} for C3 and \SI{275}{\kelvin} for C4), and then decreases at higher temperature. This behavior marks a crossover from low-temperature \gls{pte} to high-temperature \gls{nte}. The crossover becomes stronger with increasing side-chain length, demonstrating that side-chain conformational freedom is a key parameter for tuning the thermal-expansion response. 

The \gls{pimd} results in \autoref{fig:crossover}(b) confirm the same qualitative trend after including \glspl{nqe}. Although the absolute volumes and the detailed temperature dependence differ slightly from the classical \gls{md} results, the ordering of the C$n$ series and the emergence of the \gls{pte}-to-\gls{nte} crossover in the longer-chain systems remain unchanged. C0 and C2 stay in the monotonic \gls{nte} regime, whereas C3 and C4 retain the non-monotonic expansion-contraction behavior. The agreement between classical \gls{md} and \gls{pimd} indicates that the crossover is not an artifact of the classical treatment of nuclei, but a robust thermal response of the side-chain-functionalized framework. Because C5 contains the longest side chain and samples multiple metastable conformations during heating, its less regular but still recognizable \gls{pte}-to-\gls{nte} response is therefore discussed in \gls{sm} Sec.~3, while the main text focuses on C0--C4.

The microscopic picture underlying this crossover is illustrated in \autoref{fig:crossover}(c) and (d). In pristine MOF-5, transverse swinging of the primary linker shortens the projected distance between neighboring metal nodes, producing a string-tension-like contraction and giving rise to conventional \gls{nte} (\autoref{fig:crossover}(c)). In long-chain functionalized MOF-5, represented by C4, thermal side-chain opening introduces additional conformational degrees of freedom inside the pore (\autoref{fig:crossover}(d)). This side-chain-induced expansion competes with the intrinsic linker-swinging contraction of the parent framework. Thus, \autoref{fig:crossover} establishes the central physical picture: thermal expansion in the C$n$ series is governed by a competition between side-chain conformational entropy, which favors expansion, and framework vibrational entropy associated with transverse linker motion, which favors contraction.

The robustness of this picture was further tested against the treatment of dispersion interactions and force-field choice (\gls{sm} Sec.~5 and Sec.~6). Removing the D3 correction (using \gls{nep} alone) weakens the low-temperature \gls{pte} regime, especially for C3, whereas C4 retains a clear positive-to-negative crossover (\gls{sm} Figure S4). The traditional MOF-FF force field~\cite{bureekaew2013mof, pallach2021frustrated} gives a similar qualitative trend to the no-D3 \gls{nep} result, with the crossover weakened but still mainly preserved in C4 (\gls{sm} Figure S6). These comparisons show that the side-chain-length-driven crossover is qualitatively robust, while its quantitative strength depends on the accuracy with which side-chain-framework dispersion interactions and anharmonic framework dynamics are described.

The behavior in \autoref{fig:crossover} is distinct from conventional thermal-expansion responses in framework materials. Classical \gls{nte} systems, including ZrW$_2$O$_8$/HfW$_2$O$_8$ and pristine \gls{mof}-5, generally show monotonic contraction driven by transverse framework vibrations, rigid-unit modes, or linker tilting~\cite{evans1996negative,zhou2008origin,lock2010elucidating,han2007metal}. Existing \gls{mof} design strategies, such as mixed-linker solid solutions, topological design, guest adsorption, or framework retrofitting, mainly tune the magnitude or sign of thermal expansion through static structural or compositional changes~\cite{balestra2016controlling,burtch2019negative,baxter2019tuning,schneider2019tuning}. By contrast, the C$n$ series exhibits a temperature-driven crossover within an empty framework: longer side chains first promote expansion through conformational entropy and then enhance contraction through transverse framework motion. This identifies side-chain length as a molecular handle for programming a non-monotonic \gls{pte}-to-\gls{nte} response in \gls{mof}.

\begin{figure}[htb]
\begin{center}
\includegraphics[width=\columnwidth]{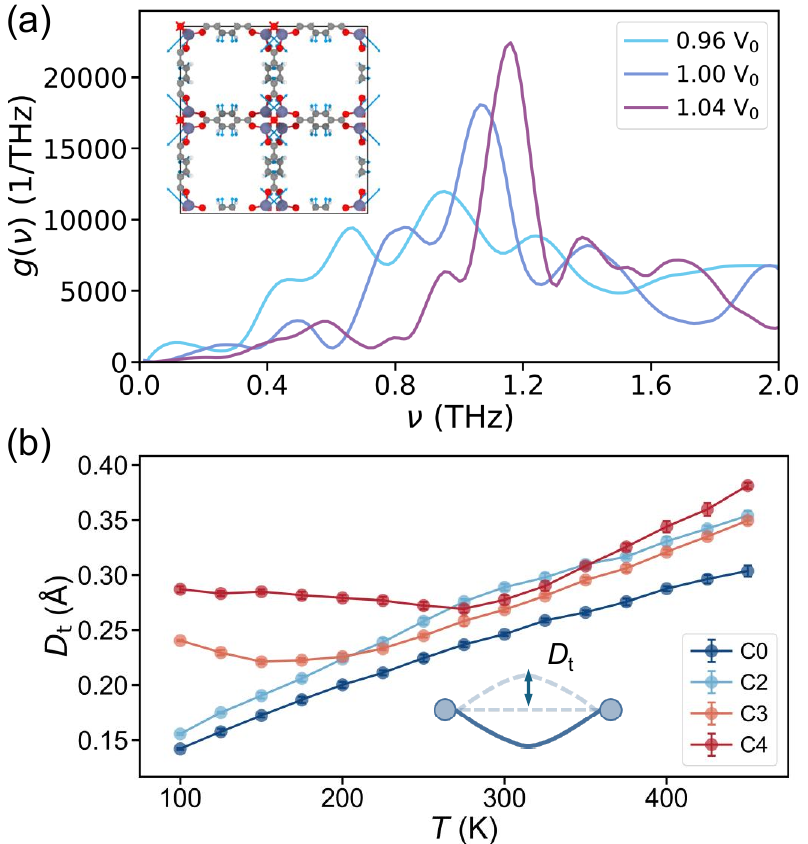}
\caption{String-tension origin of \gls{nte}.
(a) Volume-dependent low-frequency \gls{vdos} at a temperature of \SI{100}{\kelvin}, \(g(\nu)\), of the pristine C0 framework calculated for a \(5\times5\times5\) supercell at different normalized volumes. The hardening of low-frequency modes upon lattice expansion indicates their negative Gr\"uneisen character. The inset shows the \(\Gamma\)-point vibrational mode at \SI{1.059}{\tera\hertz}, with blue arrows highlighting the transverse linker motions that underlie the string-tension mechanism. (b) Temperature-dependent transverse displacement \(D_{\mathrm{t}}\) of the primary framework linkers for C0--C4. The schematic insets illustrate the lateral swinging motion of the linker and the definition of \(D_{\mathrm{t}}\).}
\label{fig:NTE}
\end{center}
\end{figure}

Having established this macroscopic crossover, we next examine the vibrational mechanism that drives framework contraction. In flexible frameworks, thermal expansion is strongly governed by low-frequency collective modes, which are readily thermally populated and can carry substantial vibrational entropy. For pristine MOF-5, the volume dependence of the low-frequency \gls{vdos} provides direct evidence for the microscopic origin of \gls{nte} (see \gls{sm} Sec.~7 for details). As shown in \autoref{fig:NTE}(a), the dominant low-frequency peaks around \SI{1}{\tera\hertz} shift toward higher frequencies when the lattice is expanded from \(0.96V_0\) to \(1.04V_0\), where \(V_0\) is the equilibrium volume of the reference pristine MOF-5 structure at a temperature of \SI{100}{\kelvin}. This anomalous hardening upon expansion corresponds to negative mode Gr\"uneisen parameters, \(\gamma_i=-\partial \ln \omega_i/\partial \ln V<0\)~\cite{gruneisen1912theorie}. Because the volumetric thermal expansion coefficient is governed by the heat-capacity-weighted average of mode Gr\"uneisen parameters, these low-frequency negative-\(\gamma_i\) modes contribute directly to \gls{nte}. The observed hardening of the low-frequency modes under lattice expansion therefore reveals an intrinsic vibrational tendency of the framework to contract upon heating.

The character of these negative-\(\gamma_i\) modes is illustrated in the inset of \autoref{fig:NTE}(a). Rather than local bond-stretching vibrations, these modes correspond to collective transverse motions of the organic linkers coupled to the Zn$_4$O nodes. Neighboring linkers swing laterally in a concerted manner, producing hinge-like framework deformations. Such transverse motion reduces the projected distance between adjacent inorganic nodes, analogous to the shortening of a vibrating string, and therefore provides the microscopic basis for the string-tension mechanism of \gls{nte}. This interpretation is broadly consistent with previous studies of \gls{nte} in \gls{mof}-5 and related framework materials~\cite{zhou2008origin,lock2010elucidating,han2007metal}.

To quantify the temperature evolution of this transverse linker motion, we calculated the transverse displacement \(D_{\mathrm{t}}\), as summarized in \autoref{fig:NTE}(b). Here, \(D_{\mathrm{t}}\) measures the lateral deviation of the primary linker from its equilibrium backbone direction and thus serves as a real-space descriptor of the string-tension mechanism. For C0 and C2, \(D_{\mathrm{t}}\) increases almost monotonically with temperature, indicating that the transverse linker fluctuations are progressively enhanced upon heating. The larger lateral displacement shortens the projected linker length along the lattice direction and pulls neighboring metal nodes closer, thereby driving framework contraction. This monotonic increase in \(D_{\mathrm{t}}\) is therefore consistent with the monotonic \gls{nte} observed in the short-chain systems.

In contrast, the longer-chain systems (C3 and C4), display a distinct temperature dependence of \(D_{\mathrm{t}}\). In the low-temperature regime, \(D_{\mathrm{t}}\) changes only weakly, indicating that transverse swinging of the primary linker is not yet strongly activated during the initial heating stage. Consequently, the string-tension-driven contraction is weak, allowing the side-chain-induced expansion to dominate. Upon further heating, however, \(D_{\mathrm{t}}\) increases rapidly, signaling the activation of stronger lateral linker fluctuations. Once this motion is activated, the attached side chains further enhance the transverse displacement of the primary linker through a mass-loading effect: the flexible side chains behave as additional weights attached to the linker, analogous to a loaded pendulum, thereby increasing the amplitude of lateral swinging. The enlarged \(D_{\mathrm{t}}\) strengthens the string-tension mechanism and drives framework contraction at high temperature. This delayed onset of transverse linker motion in C3 and C4 reflects the fact that the alkoxy chains must first thermally open before they can amplify lateral swinging; we therefore turn to the competing entropic contribution from side-chain conformational disorder, which also drives the low-temperature \gls{pte}.

\begin{figure}[htb]
\begin{center}
\includegraphics[width=\columnwidth]{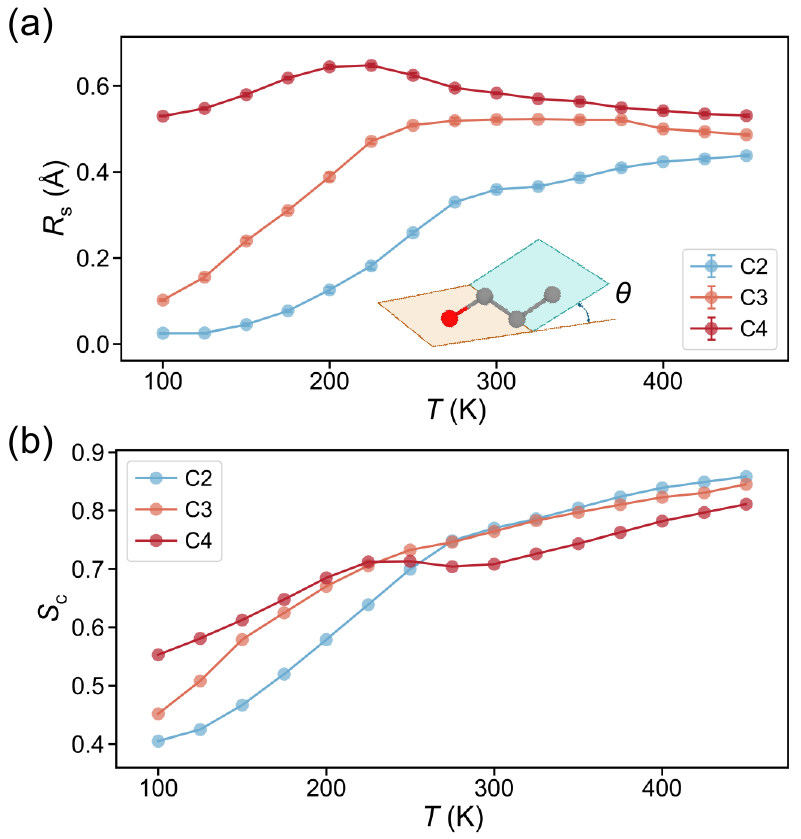}
\caption{
Conformational entropy origin of low temperature \gls{pte}. (a) The side-chain opening ratio \(R_{\mathrm{s}}\) for C2--C4 as a function of temperature; the inset schematically illustrates the opening angle \(\theta\) of the C3 structure used to define \(R_{\mathrm{s}}\). (b) Normalized configurational entropy \(S_{\mathrm{c}}\) calculated from the side-chain dihedral distributions.
}
\label{fig:PTE}
\end{center}
\end{figure}

After identifying the vibrational origin of \gls{nte}, we next examine the competing entropic contribution introduced by the flexible side chains. Unlike the transverse framework vibrations discussed above, side-chain motion mainly affects thermal expansion through conformational entropy. Upon heating, the alkoxy chains can rotate around their internal bonds and explore additional dihedral states inside the pore. This thermally activated conformational disorder increases the entropy of the side chains and generates steric pressure against the surrounding framework, thereby favoring lattice expansion.

To characterize the conformational activation of the side chains, we introduce two structural descriptors: the opening ratio \(R_{\mathrm{s}}\) and the normalized configurational entropy \(S_{\mathrm{c}}\). 
The side-chain opening angle \(\theta\) is defined from the relevant dihedral coordinates of each alkoxy chain, with the definitions for C2, C3, and C4 illustrated in \gls{sm} Figure~S7. 
The opening ratio \(R_{\mathrm{s}}\) measures the fraction of side-chain conformations whose opening angle exceeds \(60^\circ\),
\begin{equation}
R_{\mathrm{s}} =
\frac{N(\theta>60^\circ)}{N_{\mathrm{all}}},
\end{equation}
where \(\theta\) is the side-chain opening angle and \(N_{\mathrm{all}}\) is the total number of sampled conformations. 
The configurational entropy is calculated from the probability distribution of side-chain conformational states~\cite{shannon1948mathematical,karplus1981method},
\begin{equation}
S_{\mathrm{c}}(T)=
-\frac{\sum_b p_b(T)\ln p_b(T)}{\ln M},
\end{equation}
where \(p_b(T)\) is the probability of finding a side-chain conformation in bin \(b\), and \(M\) is the total number of bins in the conformational space. 
The detailed definitions of \(\theta\) for C2--C4, the treatment of the two internal dihedral angles in C4, and the binning procedure used to compute \(S_{\mathrm{c}}\) are given in the \gls{sm} Sec.~6.

As shown in \autoref{fig:PTE}(a), \(R_{\mathrm{s}}\) increases with temperature for all functionalized systems, indicating progressive thermal activation of side-chain rotations. C2 exhibits only a gradual increase, whereas C3 exhibits a rapid rise at low to intermediate temperatures, consistent with the onset of low-temperature \gls{pte}. C4 already has a large \(R_{\mathrm{s}}\) at low temperature and remains highly activated over the entire temperature range, reflecting the greater flexibility of the longer butoxy side chain.

The corresponding normalized configurational entropy \(S_{\mathrm{c}}\) is shown in \autoref{fig:PTE}(b). Compared with C3 and C4, C2 shows a smaller and more gradual increase in \(S_{\mathrm{c}}\), indicating that the shorter ethoxy side chain has fewer accessible conformational states and provides a weaker entropic driving force for expansion. For the longer-chain systems (C3 and C4), \(S_{\mathrm{c}}\) increases rapidly in the low-temperature regime, indicating that many side-chain dihedral states become thermally accessible during the initial heating stage. This rapid entropy gain provides a strong conformational driving force for lattice expansion and is therefore consistent with the low-temperature \gls{pte} observed in \autoref{fig:crossover}. At higher temperatures, however, the increase in \(S_{\mathrm{c}}\) becomes markedly slower, suggesting that many side-chain conformational states have already been sampled, and the further entropic gain per temperature increment diminishes. This allows the transverse-linker string-tension mechanism discussed above to dominate at elevated temperatures, driving the crossover from \gls{pte} to \gls{nte}.

Taken together, \autoref{fig:NTE} and \autoref{fig:PTE} reveal two competing microscopic mechanisms: transverse linker motion drives framework contraction, whereas side-chain conformational disorder favors lattice expansion. To test whether this competition can be used to program thermal expansion, we constructed mixed C0/C4 architectures with different fractions of C0- and C4-type local environments. The composition is denoted as \(w_{\mathrm{C0}}\text{@}w_{\mathrm{C4}}\), where \(w_{\mathrm{C0}}\) and \(w_{\mathrm{C4}}\) represent the molar fractions of the respective linker types. Thus, \(0.0\text{@}1.0\) corresponds to pure C4, whereas \(1.0\text{@}0.0\) corresponds to pure C0. For each intermediate composition, three independent random side-chain configurations were generated to assess configurational variability.

\begin{figure}[htb]
\begin{center}
\includegraphics[width=\columnwidth]{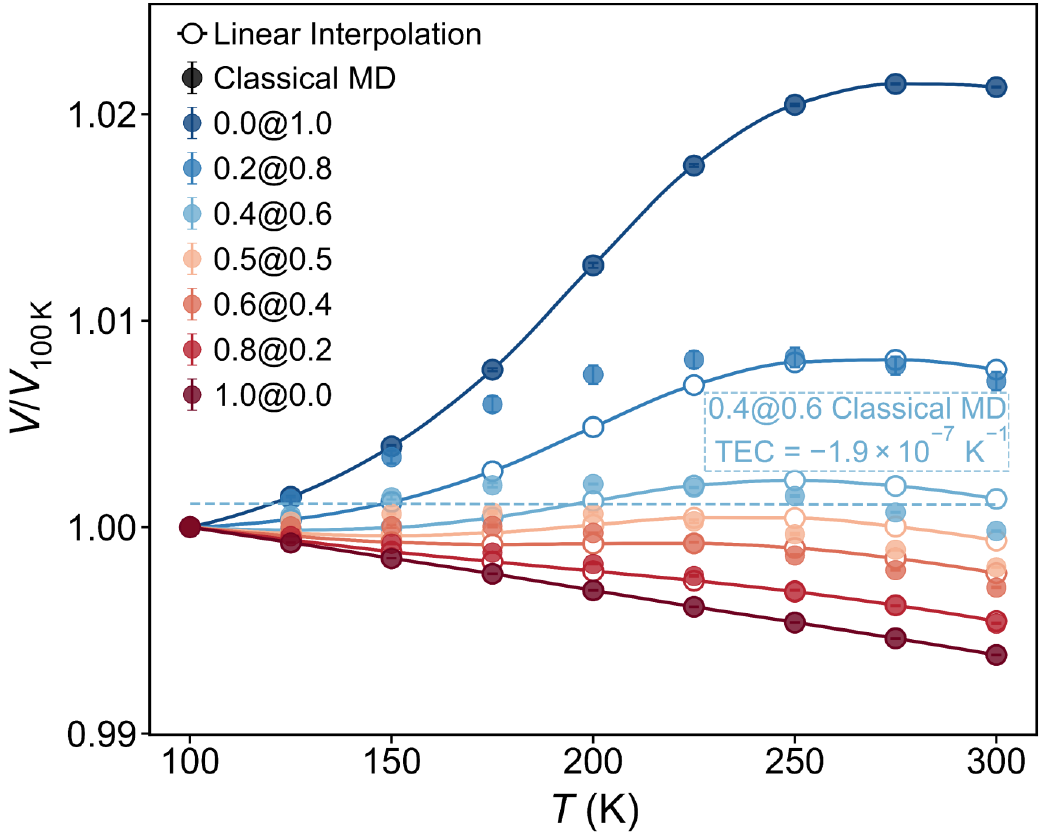}
  \caption{Composition-programmed thermal expansion over the \SIrange{100}{300}{\kelvin} range. The normalized volume \(V/V_{100\mathrm{K}}\), where \(V_{100\mathrm{K}}\) is the equilibrium volume per unit cell at \SI{100}{\kelvin}, is continuously tuned by varying the relative fraction of C0- and C4-type local environments, enabling positive, near-zero, and negative thermal expansion. Filled symbols denote classical \gls{md} results, and open symbols represent the linear interpolation.}
\label{fig:design300K}
\end{center}
\end{figure}

As shown in \autoref{fig:design300K}, the normalized volume \(V/V_{100\mathrm{K}}\) evolves continuously with composition at a temperature range of \SIrange{100}{300}{\kelvin}. C4-rich systems exhibit pronounced low-temperature expansion, whereas increasing the C0 fraction progressively suppresses this expansion and eventually yields dominant \gls{nte}. This systematic evolution demonstrates that the thermal expansion response is not limited to the discrete end-member structures, but can be continuously tuned by controlling the relative population of C0- and C4-type local environments. The corresponding behavior over a broader \SIrange{100}{450}{\kelvin} temperature range is shown in \gls{sm} Figure S8. As a simple linear predictor of this composition dependence, the volume can be approximated as $V(T) = w_{\mathrm{C0}} \lambda_{\mathrm{C0}} V_{\mathrm{C0}}(T) + w_{\mathrm{C4}} V_{\mathrm{C4}}(T)$, where \(V_{\mathrm{C0}}(T)\) and \(V_{\mathrm{C4}}(T)\) are the temperature-dependent volumes of the pure C0 and C4 systems, respectively, and \(\lambda_{\mathrm{C0}}=3.8345\) is a constant scaling factor. This linear interpolation captures the overall trend of the classical \gls{md} results. By varying the relative fraction of C0- and C4-type local environments, \(V/V_{100\mathrm{K}}\) evolves continuously from the strong low-temperature expansion of C4-rich systems to the monotonic contraction of C0-rich systems, enabling positive, near-zero, and negative \glspl{tec} within the \SIrange{100}{300}{\kelvin} range. Notably, the \(0.4\text{@}0.6\) composition exhibits nearly zero thermal expansion, with \(\alpha_V=-1.9\times10^{-7}~\mathrm{K}^{-1}\).

In summary, we have demonstrated a side-chain-programmable thermal expansion response in MOF-5. Combining classical \gls{md}, \gls{pimd}, and lattice-dynamical analyses, we show that grafting flexible alkoxy side chains onto the MOF-5 framework enables a systematic transition from monotonic \gls{nte} to a non-monotonic \gls{pte}-to-\gls{nte} crossover. Short-chain systems (C0, C2) are dominated by transverse linker vibrations and exhibit persistent \gls{nte}, whereas longer-chain systems (C3, C4) display low-temperature expansion followed by high-temperature contraction. The crossover arises from competition between two entropic mechanisms. At low temperature, thermal activation of side-chain dihedral conformations increases the configurational entropy and generates steric pressure inside the pore, driving lattice expansion. At higher temperature, the accessible conformational states become progressively exhausted, while transverse swinging of the framework linkers is strongly activated and contracts the lattice through the string-tension mechanism. For longer side chains, the alkoxy groups further amplify this transverse motion via a mass-loading effect, strengthening the high-temperature \gls{nte}. Finally, by tuning the relative fraction of C0- and C4-type local environments, the \gls{tec} can be continuously programmed, including near-zero thermal expansion at selected temperatures. Given the ease of side-chain functionalization and the vast design space of \gls{mof} linkers, this strategy should be extendable to other \glspl{mof} or soft porous crystals whose frameworks host flexible internal degrees of freedom that compete with the intrinsic vibrational response.

W. Qiu acknowledges the postdoctoral fellowships of the Sackler Center for Computational Molecular and Materials Science and the Ratner Center for Single Molecule Science at Tel Aviv University. 

\bibliography{refs}

\end{document}